\documentclass[journal=jcp,manuscript=article]{achemso}
\setkeys{acs}{maxauthors=0,articletitle=true}

\usepackage{achemso}
\usepackage{graphics}
\usepackage{amssymb,amsfonts}
\usepackage{graphicx}
\usepackage[table]{xcolor}
\usepackage{multirow}
\usepackage{caption}
\usepackage{subcaption}
\usepackage{booktabs}
\usepackage{colortbl}
\usepackage{amsmath}
\usepackage{amsopn}
\usepackage{siunitx}
\usepackage{bm}
\usepackage{color}
\usepackage{array}
\usepackage{lscape}
\usepackage{mciteplus}
\usepackage[version=3]{mhchem}
\usepackage{ulem}
\usepackage{listings}
\usepackage{enumerate}

\SectionNumbersOn

\author{Janus J. Eriksen}
\email{jeriksen@uni-mainz.de}
\affiliation[Johannes Gutenberg-Universit\"at Mainz]
{Institut f\"ur Physikalische Chemie, Johannes Gutenberg-Universit\"at Mainz, Duesbergweg 10-14, 55128 Mainz, Germany}
\author{J{\"u}rgen Gauss}
\email{gauss@uni-mainz.de}
\affiliation[Johannes Gutenberg-Universit\"at Mainz]
{Institut f\"ur Physikalische Chemie, Johannes Gutenberg-Universit\"at Mainz, Duesbergweg 10-14, 55128 Mainz, Germany}

\title[TITLE]{Many-Body Expanded Full Configuration Interaction. I. Weakly Correlated Regime}

\begin{document}

Over the course of the past few decades, the field of computational chemistry has managed to manifest itself as a key complement to more traditional lab-oriented chemistry. This is particularly true in the wake of the recent renaissance of full configuration interaction (FCI)-level methodologies, albeit only if these can prove themselves sufficiently robust and versatile to be routinely applied to a variety of chemical problems of interest. In the present series of works, performance and feature enhancements of one such avenue towards FCI-level results for medium to large one-electron basis sets, the recently introduced many-body expanded full configuration interaction (MBE-FCI) formalism [{\it{J. Phys. Chem. Lett.}} {\bf{8}}, 4633 (2017)], will be presented. Specifically, in this opening part of the series, the capabilities of the MBE-FCI method in producing near-exact ground state energies for weakly correlated molecules of any spin multiplicity will be demonstrated.

\newpage

%
%%%%%%%%%%%%%%%%%%%%%%%%%%%%%%%%%%%%%%%%%%%%%%%%%%%%%%%%%%%%%%%%%%%
%                                                                    			INTRODUCTION
%%%%%%%%%%%%%%%%%%%%%%%%%%%%%%%%%%%%%%%%%%%%%%%%%%%%%%%%%%%%%%%%%%%
%

\section{Introduction}\label{intro_section}

Owing to the fact that all facets of chemistry are ultimately governed by the set of laws that define quantum mechanics, physical chemical quantities such as, for instance, reaction energies, molecular properties, equilibrium structures, etc., may in principle be decoded directly from theory. Nevertheless, the key equation in the field of quantum chemistry, the time-independent Schr\"odinger equation, $\hat{H}|\Psi\rangle = E|\Psi\rangle$, becomes too complex to be solved analytically for any system involving more than one electron, despite its deceptively innocent appearance. This intractability is ultimately related to the fact that the Hamiltonian, $\hat{H}$, involves the repulsive Coulomb interaction among all the electrons of the system at hand. For this reason, approximations to the electronic wave function, $|\Psi\rangle$, are hence requisites needed in order to extract any kind of information from the Schr\"odinger equation.

Within a given basis of one-electron atomic orbitals (AOs), the exact solution to the Schr{\"o}dinger equation is formulated as a superposition of every possible determinant that may be generated by distributing all $N$ electrons of the system among all the molecular orbitals (MOs) from a preceding Hartree-Fock (HF) calculation (of which there is a total of $M$).
This solution is known as the full configuration interaction (FCI) wave function, the practical realization of which, its formal attractiveness aside, is generally impeded by a two-fold curse of dimensionality~\cite{knowles_handy_fci_cpl_1984,olsen_fci_jcp_1988,olsen_fci_cpl_1990}. Within a basis set of a certain quality, the scaling of the FCI model is exponential with respect to the number of electrons, and even for a fixed system size, the computational requirements grow exponentially with respect to the number of MOs. For this reason, many branches of the quantum chemistry community have over the years been devoted to approximating the FCI wave function, traditionally by limiting the space of possible determinants that may be reached from the HF determinant.

Now, rather than attempting to describe the FCI wave function from a reduced number of variable parameters, as in, for instance, coupled cluster (CC) theory~\cite{cizek_1,cizek_2,paldus_cizek_shavitt} or density matrix renormalization group~\cite{white_dmrg_prl_1992,white_dmrg_prb_1993} (DMRG) theory and its application to quantum chemistry~\cite{white_martin_dmrg_jcp_1999,mitrushenkov_palmieri_dmrg_jcp_2001,chan_head_gordon_dmrg_jcp_2002,legeza_hess_dmrg_prb_2003,chan_dmrg_2011,wouters_dmrg_2014,knecht_dmrg_2016}, an alternative idea was put forward by Malrieu in the 1970s~\cite{malrieu_selected_ci_jcp_1973} who proposed to instead describe its most {\it{important}} components in a direct manner without sacrificing the accuracy. Within the last decade, this particular philosophy has been revisited by numerous groups. In particular, the advent of powerful approaches deriving from stochastic Monte-Carlo solutions to the Schr\"odinger equation~\cite{booth_alavi_fciqmc_jcp_2009,cleland_booth_alavi_jcp_2010,booth_alavi_fciqmc_jcp_2010,booth_alavi_fciqmc_jcp_2011,cleland_booth_alavi_fciqmc_jctc_2012,daday_booth_alavi_filippi_fciqmc_jctc_2012,booth_alavi_fciqmc_nature_2013,booth_alavi_fciqmc_jcp_2017,petruzielo_umrigar_spmc_prl_2012,holmes_umrigar_heat_bath_fock_space_jctc_2016} has propagated through the community, enabling unprecedented system sizes to be examined under the lens of the computational microscope. In recent years, also proposals for new focused CI treatments have been prolific in the literature~\cite{neese_selected_ci_talk_2016,head_gordon_whaley_selected_ci_jcp_2016,holmes_umrigar_heat_bath_ci_jctc_2016,sharma_umrigar_heat_bath_ci_jctc_2017,holmes_sharma_heat_bath_ci_excited_states_jcp_2017,chien_zimmerman_heat_bath_ci_excited_states_jpca_2017,schriber_evangelista_selected_ci_jcp_2016,zhang_evangelista_projector_ci_jctc_2016,schriber_evangelista_adaptive_ci_jctc_2017,liu_hoffmann_ici_jctc_2016,fales_koch_martinez_rrfci_jctc_2018}. The common denominator in all of these is the tight connection to the original ideas of {\it{selected}}, rather than full CI. While capable of treating weak (dynamic) as well as strong (static) correlation phenomena on an unbiased, equal footing, they all, alongside DMRG and stochastic analogues, suffer from the exponential scaling discussed above, although at a much reduced prefactor and for DMRG only when applied to systems that span more than a single dimension. In the present series of works, however, attention will be given to yet another approach, one in which the FCI energy is targeted directly by means of the method of increments, thus circumventing the quest for the complex $N$-dimensional wave function altogether.

In a series of works published during the late 1960s~\cite{nesbet_phys_rev_1967_1,nesbet_phys_rev_1967_2,nesbet_phys_rev_1968}, Nesbet proposed to decompose the FCI correlation energy, $E_{\text{FCI}}$, through a many-body expansion (MBE) in the $M_{\text{o}}$ occupied spatial MOs of a system (conventionally labeled by indices $i,j,k,\ldots$)
\begin{align}
E_{\text{FCI}} &= \sum_{i}\epsilon_{i} + \sum_{i>j}\Delta\epsilon_{ij} + \sum_{i>j>k}\Delta\epsilon_{ijk} + \ldots \nonumber \\
&\equiv E^{(1)} + E^{(2)} + E^{(3)} + \ldots + E^{(M_{\text{o}})} \ . \label{bethe_goldstone_eq}
\end{align}
In Eq. \ref{bethe_goldstone_eq}, $\epsilon_{i}$ denotes the energy obtained by performing a calculation in the composite space of occupied orbital $i$ and the complete set of virtuals, while $\Delta \epsilon_{ij}$, $\Delta \epsilon_{ijk}$, etc., denote changes in electron correlation (increments) from correlating the electrons of two occupied orbitals over one, three over two, etc.~\cite{sinanoglu_jcp_1962,harris_monkhorst_freeman_1992}
\begin{subequations}
\label{increments_eqs}
\begin{align}
\Delta \epsilon_{ij} &= \epsilon_{ij} - (\epsilon_{i} + \epsilon_{j}) \label{two_body_inc} \\
\Delta \epsilon_{ijk} &= \epsilon_{ijk} - (\Delta \epsilon_{ij} + \Delta \epsilon_{ik} + \Delta \epsilon_{jk}) - (\epsilon_{i} + \epsilon_{j} + \epsilon_{k}) \ . \label{three_body_inc}
\end{align}
\end{subequations}
Due to its universality, Eq. \ref{bethe_goldstone_eq} is by no means restricted to a basis of individual MOs and one might instead choose upon domain clusters of orbitals, or molecular moieties of a supersystem even, as the objects entering the expansion. As a result, general MBE-based incremental schemes in any of these bases have been very much in vogue as of late~\cite{stoll_cpl_1992,stoll_phys_rev_b_1992,stoll_jcp_1992,friedrich_jcp_2007,xantheas_mbe_int_energy_jcp_1994,paulus_stoll_phys_rev_b_2004,stoll_paulus_fulde_jcp_2005,truhlar_mbe_jctc_2007,truhlar_mbe_dipole_mom_pccp_2012,crawford_mbe_optical_rot_tca_2014,gordon_slipchenko_chem_rev_2012,bytautas_ruedenberg_ceeis_jcp_2004,bytautas_ruedenberg_ceeis_jpca_2010,ruedenberg_windus_mbe_jpca_2017,parkhill_mbe_neural_network_jcp_2017,zgid_gseet_jpcl_2017}. The rationale behind this notable rise in popularity is the fact that Eq. \ref{bethe_goldstone_eq} becomes of practical value if contributions from higher-order increments turn out to be negligible. Returning to the case of individual MOs as the expansion objects, the FCI limit may hence be approached---at least in principle---by correlating an increasing number of electrons independently and in succession at a reduced overall cost. That being said, whereas the number of electrons may be limited, extended basis sets are compulsory for solving the Schr\"odinger equation, and the FCI dimensionality curse will still prevail at high orders, much alike the scenario in any of the FCI-level approximations discussed above.

In a recent letter~\cite{eriksen_mbe_fci_jpcl_2017}, the present authors instead proposed to turn things around by considering the objects of the MBE not to be the occupied, but rather the virtual MOs of the system, in an algorithm referred to as many-body expanded FCI (MBE-FCI). This way, the number of independent calculations will increase upon moving to larger basis sets (as opposed to Eq. \ref{bethe_goldstone_eq}), while the cost of the individual calculations remains marginal, operating under the assumption that Eq. \ref{bethe_goldstone_eq} still converges reasonably fast. Facilitated by a screening protocol for eliminating the well-known redundancy in the FCI wave function~\cite{knowles_handy_fci_jcp_1989,ivanic_ruedenberg_ci_deadwood_tca_2001,bytautas_ruedenberg_ci_deadwood_cp_2009} and further aided by expansion starting points different from the HF solution, such as that of the coupled cluster singles and doubles~\cite{ccsd_paper_1_jcp_1982} (CCSD) solution, results to within sub-kJ/mol accuracy were reported in Ref. \citenum{eriksen_mbe_fci_jpcl_2017} for the prototypical H$_2$O system in core-valence basis sets ranging from double- to quadruple-$\zeta$ quality. Despite having been obtained using a pilot implementation of the algorithm, these initial results were still testament to the fact that the fundamental idea behind MBE-FCI indeed works as intended without recourse to an explicit truncation of the orbital space extent in the individual increment calculations, as is the case in the incremental FCI (iFCI) algorithm by Zimmerman~\cite{zimmerman_ifci_jcp_2017_1,zimmerman_ifci_jcp_2017_2,zimmerman_ifci_jpca_2017}, which is another recent incarnation of an MBE-based approach to the FCI problem, albeit one operating in the traditional basis of occupied orbitals.\\

In the present series of works, we will introduce a number of improvements made to the computational potential of the MBE-FCI method, as presented in Ref. \citenum{eriksen_mbe_fci_jpcl_2017}, in addition to a number of comprehensive enhancements of core functionality made to enlarge its general application range. Whereas focus will once again be on small molecular systems dominated by weak correlation in this first part of the series, forthcoming parts will be devoted to strongly correlated systems, ground and excited state first-order properties, as well as systems of larger overall size, all in the context of MBE-FCI.

%
%%%%%%%%%%%%%%%%%%%%%%%%%%%%%%%%%%%%%%%%%%%%%%%%%%%%%%%%%%%%%%%%%%%
%                                                                    				THEORY
%%%%%%%%%%%%%%%%%%%%%%%%%%%%%%%%%%%%%%%%%%%%%%%%%%%%%%%%%%%%%%%%%%%
%

\section{Theory}\label{theory_section}

The master equation in MBE-FCI is the decomposition of the FCI correlation energy in Eq. \ref{bethe_goldstone_eq}, but formulated in terms of virtual rather than occupied MOs (labeled by indices $a,b,c,\ldots$)
\begin{align}
E_{\text{FCI}} &= \sum_{a}\epsilon_{a} + \sum_{a>b}\Delta\epsilon_{ab} + \sum_{a>b>c}\Delta\epsilon_{abc} + \ldots \nonumber \\
&\equiv E^{(1)} + E^{(2)} + E^{(3)} + \ldots + E^{(M_{\text{v}})} \label{mbe_eq}
\end{align}
where $M_{\text{v}}$ designates the number of virtual MOs in the system. Hence, in analogy with the original formalism, the increments now account for changes in electron correlation from allowing for electronic excitations into an increasing number of unoccupied orbitals. 

\subsection{Screening}\label{mbe_fci_subsection}

At this point, it will prove instructive to revisit the physical reasoning behind any MBE; regardless of the involved expansion objects, be they occupied or virtual orbitals or even domains of any of these, MBE-based methods approximate $n$-body effects by means of additive cumulants
\begin{align}
\epsilon_{[\Omega]} = \sum_{p \in S_1[\Omega]_{n}}\epsilon_{p} + \sum_{pq \in S_2[\Omega]_{n}}\Delta\epsilon_{pq} + \ldots + \Delta\epsilon_{[\Omega]} \label{cumulant_eq}
\end{align}
where the action of $S_{m}$ onto a tuple of $n$ orbitals, $[\Omega]_{n} \equiv [ab\cdots]_{n}$, is to construct all possible unique subtuples of order (length) $m$ where $1\leq m<n$. Under standard conditions, that is, in the absence of strong correlation, the magnitude of individual increments will decrease upon moving to higher orders in the expansions, but these cannot be guaranteed to take any particular sign ($\pm$). The immediate implication of this conundrum is thus that one cannot safely establish convergence on the basis of a simple energy-based criterion (e.g., difference between two successive orders) as this is prone to false convergences in the sense that increments may cancel when these are of relatively large magnitude, but alternating signs. At the same time, an accumulation of modest-sized, same-sign increments may result in sizable order corrections, and these are bound to be missed from any algorithm that solely monitors a change in the total energy and terminates whenever this falls below some threshold. However, this claim only touches upon the general lack of monotonicity in any MBE, not the fact that it ultimately converges. Indeed, in Ref. \citenum{eriksen_mbe_fci_jpcl_2017} it was shown that oscillatory converging behaviour is the rule rather than the exception in MBE-FCI---even for simple weakly correlated systems---a point which will be numerically reiterated in Section \ref{results_section}, but this does not imply that one is necessarily forced to calculate every single of the copious body of increments that appear at increasingly higher orders in the MBE. Rather, the redundancy associated with calculating zeros may be avoided through the introduction of a screening protocol that aims at filtering out numerically vanishing increments up through the expansion. We will now provide a brief summary of how screening is performed in the MBE-FCI algorithm.

At each order $k$, all possible {\it{child tuples}} at the following order $k+1$ are generated from the complete set of $k$-order {\it{parent tuples}} subject to an expansion threshold. Hence, the protocol seeks to predict the relative magnitude of higher-order increments in order not to explicitly evaluate them. Specifically, for each parent tuple at order $k$, denoted as $[ab\cdots c]_{k}$, we probe whether or not to consider a distinct child tuple, $[ab\cdots cd]_{k+1}$, at order $k+1$ by initially constructing the following set of tuples of length $k$
\begin{align}
\{\Lambda\}_{k} &= S_{k-1}[ab\cdots c]_{k} \otimes [d]_{1} \label{lambda_set}
\end{align}
where the action of $S_{k-1}$ is the same as in Eq. \ref{cumulant_eq} and the direct product produces all combinations that append the MO $d$ to any of the unique subtuples of length $k-1$. The fate of the child tuple in question, $[ab\cdots cd]_{k+1}$, is now governed by the following screening condition
\begin{align}
T_{k} &< \max_{\lambda \in \{\Lambda\}_{k}}|\Delta\epsilon_{\lambda}| \label{screen_cond}
\end{align}
for some numerical energy threshold, $T_{k}$, see below. That is, if the orbital $d$ is estimated as being sufficiently correlated with {\it{any}} combination of orbitals present in the parent tuple, then said child tuple will be among the tuples considered at order $k+1$, and vice versa, if the condition in Eq. \ref{screen_cond} fails to be satisfied. In the original version of the protocol~\cite{eriksen_mbe_fci_jpcl_2017}, the orbital $d$ was required to be sufficiently correlated with {\it{all}} combinations of orbitals present in the parent tuple, that is, the condition for generating the child tuple in question was
\begin{align}
T_{k} &< \min_{\lambda \in \{\Lambda\}_{k}}|\Delta\epsilon_{\lambda}| \label{screen_cond_old}
\end{align}
instead of the current condition in Eq. \ref{screen_cond}. Since the implications of screening implicitly propagate to higher orders, as no potential higher-order tuples may in turn be inherited from $[ab\cdots cd]_{k+1}$, the protocol guarantees that the MBE terminates at some order $k \leq M_{\text{v}}$ (the equality sign holding in the case of $T_{k} = 0.0 \ E_{\text{H}} \ \forall \ k \in [1,M_{\text{v}}-1]$). The tighter the threshold, the more computationally demanding the expansion thus becomes. However, while the graph-like generation of input tuples necessitates a tight threshold early on in the expansion, this is less decisive upon moving to higher orders if indeed all of the increments belonging to the increasingly large manifolds become increasingly negligible---in the present work, we will monitor exactly this by recording the absolute magnitude of the largest increment at each order (see Section \ref{results_section}). For instance, we might opt to entirely avoid screening at the first couple of orders followed by screening subject to a finite numerical threshold, which may even be relaxed along the path of the expansion
\begin{align}
T_{k} \ (\text{in} \ E_{\text{H}}) \equiv 
\left\{\begin{array}{lr}
0.0 & \text{if } k < k_{\text{init}} \ \phantom{.} \\
T_{\text{init}} \cdot a^{k-k_{\text{init}}} & \text{if } k \geq k_{\text{init}} \ .
\end{array}\right. \label{threshold}
\end{align}
For all of the calculations to follow in Section \ref{results_section}, the relaxation factor, $a$, initial screening order, $k_{\text{init}}$, and initial threshold, $T_{\text{init}}$, are chosen as $a \geq 1.0$, $k_{\text{init}} = 3$, and $T_{\text{init}} = \num{1.0e-10}$ $E_{\text{H}}$, respectively, in practice leaving $a$ as the only adjustable MBE-FCI parameter. Furthermore, $T_{\text{init}}$ marks the value to within which the energies of the individual complete active space CI (CASCI) and CC calculations are converged, and hence a conservative lowest threshold for which the numerical precision of the calculation may be controlled~\cite{herbert_mbe_jcp_2014,herbert_mbe_acc_chem_res_2014}. With respect to the original version of the protocol outlined in Ref. \citenum{eriksen_mbe_fci_jpcl_2017}, the current version presented above is thus made slightly less aggressive, although significantly more safe and robust (cf. Section \ref{screening_subsection} where the use of Eq. \ref{screen_cond} over Eq. \ref{screen_cond_old} is evaluated), but at the same time the additional cost associated with these changes has been compensated for by technical enhancements made to the underlying code, cf. Section \ref{comp_section}. 

\subsection{Recent Advancements}\label{new_subsection}

Besides the use of screening, the MBE-FCI algorithm offers yet another means to accelerate convergence, namely the freedom to choose an arbitrary non-HF {\it{base}} for the expansion. More precisely, rather than the entire FCI correlation energy, the expansion may target the gap in correlation energy between the solution for the full system obtained at some intermediate level, $x$, and FCI
\begin{align}
E_{\text{FCI}} &= E_{x} + \sum_{a}(\epsilon_{a,\text{FCI}}-\epsilon_{a,x}) + \sum_{a>b}(\Delta\epsilon_{ab,\text{FCI}}-\Delta\epsilon_{ab,x}) + \ldots \nonumber \\
&= E_{x} + \sum_{a}\tilde{\epsilon}_{a} + \sum_{a>b}\Delta\tilde{\epsilon}_{ab} + \ldots \label{mbe_base_eq}
\end{align}
While the use of Eq. \ref{mbe_base_eq} assumes that a calculation at this lower level of theory can be performed for the full system prior to commencing the actual expansion, as well as within each of the CAS spaces of the individual tuple calculations, the clear advantage of using such an intermediate model is that the individual energy increments are bound to be notably smaller in value, leading to a potentially faster convergence towards the FCI solution. However, as we will highlight through the results in Sections \ref{screening_subsection}, the convergence rate of Eq. \ref{mbe_base_eq} relies not only on that of the FCI expansion in Eq. \ref{mbe_eq}, but also the premise that the convergence rates for both of the involved models are comparatively similar. If not, spurious artefacts may taint the overall expansion, in particular in the absence of screening.

\begin{figure}[ht]
\begin{center}
\includegraphics[width=\textwidth]{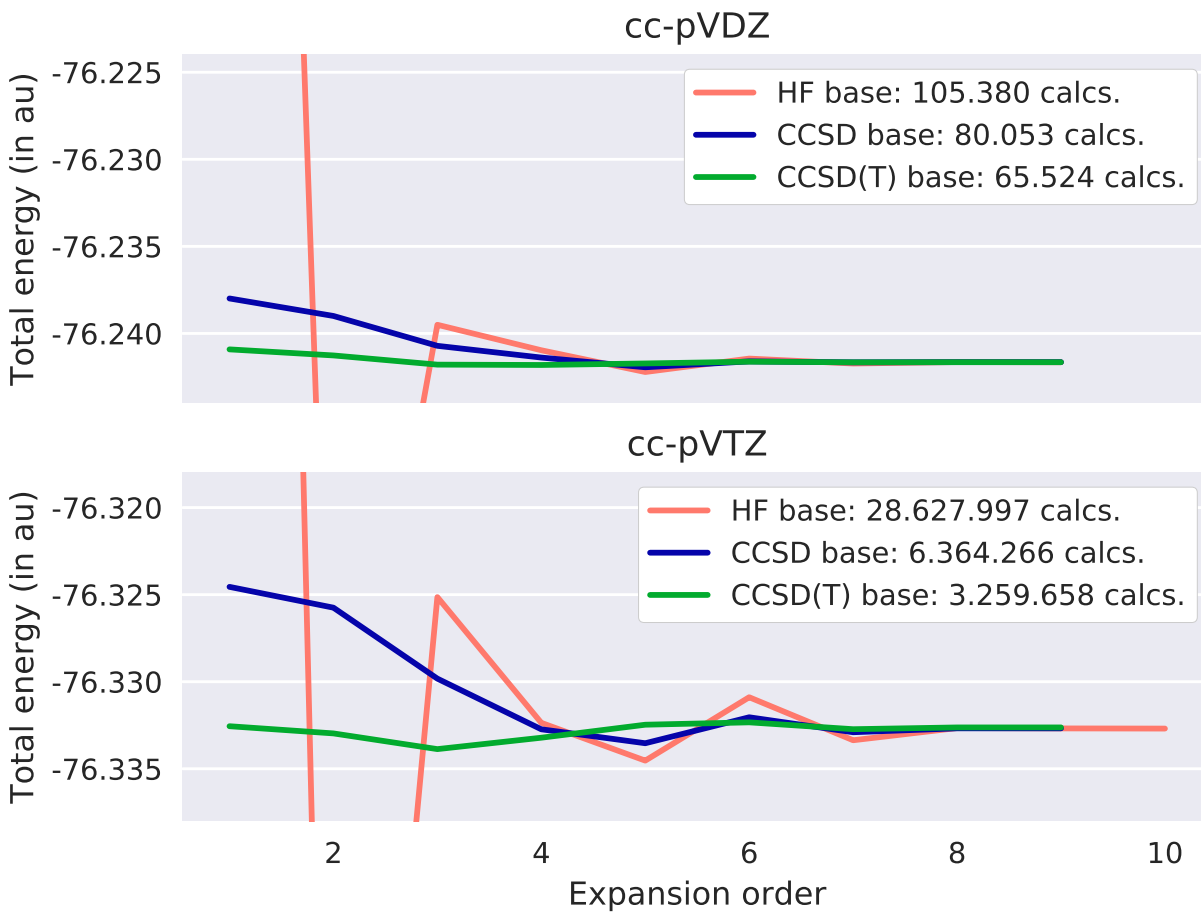}
\caption{Energy convergence and corresponding number of calculations involved in MBE-FCI expansions with various base models for the ${^{1}}\text{A}_1$ ground state of H$_2$O in the cc-pVDZ ($8$e,$23$o) and cc-pVTZ ($8$e,$57$o) basis sets. In all calculations, a screening threshold of $a = 5.0$ was used.}
\label{base_model_fig}
\end{center}
\end{figure}
In Ref. \citenum{eriksen_mbe_fci_jpcl_2017}, CCSD was successfully employed as a base model and recently the option to also use the CCSD(T) model~\cite{original_ccsdpt_paper,ccsdpt_perturbation_stanton_cpl_1997} has been added to the code. As is illustrated in Figure \ref{base_model_fig}, the favourable compromise between accuracy and cost in the CCSD(T) model~\cite{triples_pert_theory_jcp_2015} may aid in accelerating convergence even further. In addition, one may opt to diagonalize the virtual-virtual block of the one-particle reduced density matrix at the base level of theory in order to obtain a set of virtual natural orbitals (NOs), whenever an energy calculation at this level of theory is indeed possible for the full system. In turn, the use of NOs allows for more accelerated expansions over the use of standard canonical virtual HF orbitals~\cite{lowdin_nat_orb_fci_phys_rev_1955,lowdin_nat_orb_fci_phys_rev_1956}. In generating the results of Figure \ref{base_model_fig} as well as those to follow in Section \ref{results_section}, CCSD virtual NOs and canonical occupied MOs were used regardless of the chosen base model. However, we note how this particular choice of orbital representation has implications in the case of a CCSD(T) base model, as the diagonal part of the virtual-virtual block of the Fock matrix, $f_{aa}$, is now substituted in place of virtual MO energies, both in the preceding base calculation as well as in the individual CASCI calculations. While this substitution is strictly speaking in violation with the standard definition of the CCSD(T) model~\cite{original_ccsdpt_paper}, we are, in the present context, not necessarily concerned with the formal correctness of any base model, but rather its ability to mirror the incremental correlation of FCI throughout an MBE. This is an important feature of every choice of base model in MBE-FCI, as the approximated FCI energy remains the only quantity of interest.

Finally, the option to use restricted open-shell HF (ROHF) reference determinants for open-shell species has recently been implemented by treating singly and doubly occupied orbitals on an equal footing as members of the total set of occupied MOs. Stated differently, singly occupied orbitals are excluded from the virtual expansion set, and for any molecule which may exist in isoelectronic closed- and open-shell configurations, the number of virtual orbitals will hence differ. Besides this, no other significant modifications are necessary, as long as a proper implementation of the MBE-FCI algorithm takes care in monitoring for correct spin multiplicity and symmetry in all of the individual subcalculations. However, we note that in the treatment of open-shell species by post-HF methods, an ROHF trial function has often been preferred over a corresponding unrestricted (UHF) reference whenever the expectation value $\langle \hat{S}^{2} \rangle$ for the latter type deviates significantly from the correct value. Both types, however, offer valid choices for the uncorrelated treatment of open-shell systems in the sense that both have pros and cons to their general use~\bibnote{Whereas a semicanonical reference is conventionally required when CCSD(T) is employed on top of ROHF references~\cite{rittby_bartlett_rohf_ccsd_jpc_1988,watts_rohf_ccsdpt_jcp_1993}, the pragmatism exerted in case of RHF references, for which virtual MO energies are approximated by diagonal Fock elements, is repeated also for open-shell cases. Furthermore, the presence of off-diagonal Fock matrix elements ($f_{ia}$ and $f_{ai}$) is ignored in all ROHF-CCSD(T) base model calculations. The argument from earlier is, however, reiterated, as the quantitative performance of the resulting CCSD(T)-like base model remains preserved irrespective of the loss of theoretical exactness, making it perfectly suitable for its present purposes as an MBE-FCI base model.}.

%
%%%%%%%%%%%%%%%%%%%%%%%%%%%%%%%%%%%%%%%%%%%%%%%%%%%%%%%%%%%%%%%%%%%
%                                                                    		COMPUTATIONAL DETAILS
%%%%%%%%%%%%%%%%%%%%%%%%%%%%%%%%%%%%%%%%%%%%%%%%%%%%%%%%%%%%%%%%%%%
%

\section{Computational Details}\label{comp_section}

All results to follow in Section \ref{results_section} have been obtained using the new {\textsc{pymbe}} code~\cite{pymbe}, which is written in Python/NumPy~\cite{numpy} and utilizes the {\textsc{pyscf}} program~\cite{pyscf_prog,pyscf_paper} for all electronic structure kernels. The embarrassingly parallel nature of the entire MBE-FCI algorithm has been exposed by means of the message passing interface (MPI) standard via its implementation in the {\sc{mpi4py}} Python module~\cite{mpi4py_1,mpi4py_2,mpi4py_3}, resulting in massive parallelism at the expense of a bare minimum of interprocess communication. As opposed to the original pilot implementation, the {\textsc{pymbe}} code now supports full Abelian point-group symmetry---with the direct implication that base model NOs become symmetry adapted---and all energy summations of the algorithm are now performed in an exactly rounded manner via the Python {\texttt{fsum}} function, which implements the Shewchuk algorithm to track multiple partial sums~\cite{shewchuk_sum_1997}. Finally, a number of technical improvements to both the {\textsc{pymbe}} and the underlying {\textsc{pyscf}} codes have been made. In the former, an MPI bottleneck, related to a computational overhead in the handling of two-electron repulsion integrals on the slaves processes, has been resolved, and the vectorization of the screening protocol has been considerably improved, as has the load balancing in all MPI phases of the code, which themselves have been formulated in terms of non-blocking communication patterns. Within {\textsc{pyscf}}, the option to perform MO-based CC calculations fully in-core has been enabled, such that these are run on par with all CASCI calculations (i.e., free of any I/O). Accordingly, the current generation of {\textsc{pymbe}} shows significantly improved time-to-solution and strong scalability, even when calculations are confined to a single node and the parallel execution takes place among the individual cores local to this.

In terms of computer hardware, all calculations have been performed on a single, admittedly reasonably large node comprising a total of 44 cores {@} 2.20 GHz distributed over 2 Intel Xeon Broadwell E5-2699 v4 CPUs and $768$ GB of global memory. Despite the fact that the MBE-FCI algorithm is overall amenable to large-scale applications, the use of single-node commodity hardware provides us with an opportunity to challenge the assumption that large computer facilities are necessary for such formalisms to be useful. In practical terms, however, for general calculations in extended basis sets to become feasible (timing-wise), a proper massively parallel setup must be secured. For this reason, scalability demonstrations as well as calculations in basis sets larger than quadruple-$\zeta$ quality are postponed for future studies. We stress, though, that calculations within large basis sets are indeed tractable, as was previously highlighted in Ref. \citenum{eriksen_mbe_fci_jpcl_2017}, where the convergence behavior of MBE-FCI onto the FCI target was shown to be invariant upon an increase in basis sets size.

%
%%%%%%%%%%%%%%%%%%%%%%%%%%%%%%%%%%%%%%%%%%%%%%%%%%%%%%%%%%%%%%%%%%%
%                                                                    				RESULTS
%%%%%%%%%%%%%%%%%%%%%%%%%%%%%%%%%%%%%%%%%%%%%%%%%%%%%%%%%%%%%%%%%%%
%

\section{Results}\label{results_section}

In the present Section, we will operate with three distinct expansion types, all using identical representations for the occupied (HF canonical) and virtual (CCSD natural) orbital spaces while differing in the involved base model; for the sake of brevity, we introduce a shorthand notation to denote each separate expansion, namely by identifying them by their base model---{\textbf{b:None}}, {\textbf{b:CCSD}}, and {\textbf{b:CCSD(T)}}. The frozen-core approximation is invoked throughout unless otherwise stated, and all comparative FCI (where applicable), semistochastic heat-bath CI~\cite{holmes_umrigar_heat_bath_fock_space_jctc_2016,sharma_umrigar_heat_bath_ci_jctc_2017,holmes_sharma_heat_bath_ci_excited_states_jcp_2017} (SHCI), and CC with up to quadruple excitations~\cite{ccsdtq_paper_1_jcp_1991,ccsdtq_paper_2_jcp_1992} (CCSDTQ) calculations of the present work have been calculated using the {\textsc{pyscf}}, {\textsc{dice}}~\cite{dice,smith_sharma_heat_bath_casscf_jctc_2017}, and {\textsc{mrcc}}~\cite{mrcc,kallay_string_based_cc_jcp_2001} codes, respectively, the latter two through their interfaces in the {\textsc{pyscf}} and {\textsc{cfour}}~\cite{cfour} codes, respectively. All results are reported as correlation energies or differences between these, and all are defined with respect to RHF and ROHF reference energies for closed- and open-shell species, respectively.

\subsection{Screening Revisited}\label{screening_subsection}

Before turning our attention to other molecular species, we initially return to the case of water which was studied in Ref. \citenum{eriksen_mbe_fci_jpcl_2017}. In the scope of the present work, we will use H$_2$O as a testbed to validate the improvements made to the screening protocol, that is, Eq. \ref{screen_cond} over Eq. \ref{screen_cond_old}. However, unlike in Ref. \citenum{eriksen_mbe_fci_jpcl_2017}, we will this time use the geometry of Ref. \citenum{zimmerman_ifci_jcp_2017_1} while employing the standard cc-pV$X$Z basis sets~\cite{dunning_1_orig} in order to facilitate numerical comparisons to some of the alternative FCI-like methods discussed in Section \ref{comparison_subsection}.

\begin{figure}[ht]
\begin{center}
\includegraphics[width=\textwidth]{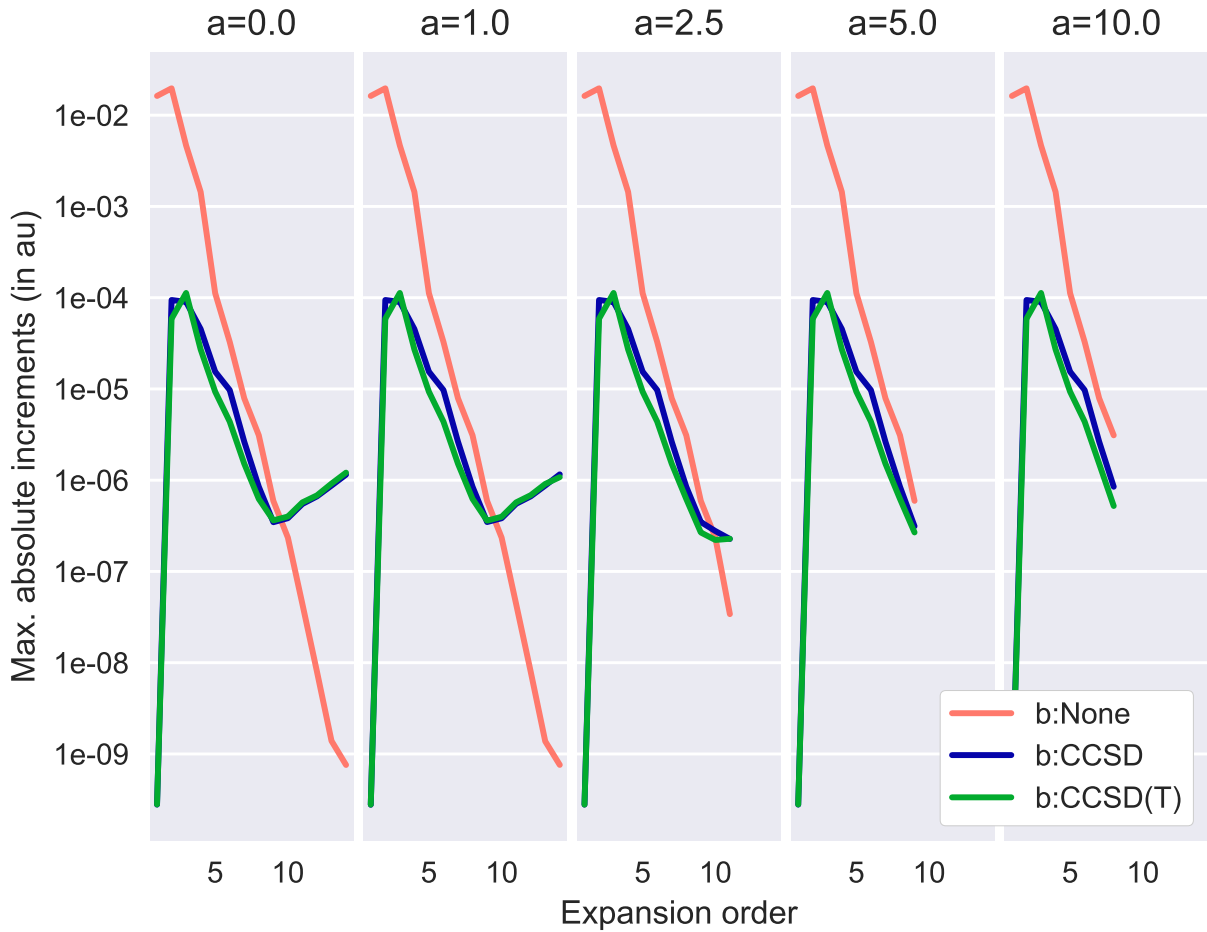}
\caption{Evolution of the largest absolute increment in any of the MBE-FCI/cc-pVDZ expansions of Table \ref{screening_table} for the ${^{1}}\text{A}_1$ ground state of H$_2$O ($8$e,$23$o).}
\label{inc_max_h2o_dz_relax_fig}
\end{center}
\end{figure}

\begin{table}[ht]
\begin{center}
\begin{tabular}{l|ccc|c}
\toprule
\multicolumn{1}{c|}{Threshold} & \multicolumn{1}{c}{{\textbf{b:None}}} & \multicolumn{1}{c}{{\textbf{b:CCSD}}} & \multicolumn{1}{c}{{\textbf{b:CCSD(T)}}} & \multicolumn{1}{c}{$\mu\pm\sigma$} \\
\midrule\midrule
$a = 10.0$ & $-214.82$ & $-214.81$ & $-214.80$ & $-214.81\pm0.02$ \\
$a = 5.0$ & $-214.80$ & $-214.79$ & $-214.80$ & $-214.80\pm0.01$ \\
$a = 2.5$ & $-214.80$ & $-214.80$ & $-214.80$ & $-214.80\pm0.00$ \\
$a = 1.0$ & $-214.80$ & $-214.83{^{\ast}}$ & $-214.82{^{\ast}}$ & $-214.82\pm0.04$ \\
$a = 0.0$ & $-214.80{^{\ast}}$ & $-214.93{^{\ast}}$ & $-214.94{^{\ast}}$ & $-214.89\pm0.19$ \\
\midrule
\multicolumn{1}{c|}{$\mu\pm\sigma$} & $-214.80\pm0.01$ & $-214.83\pm0.07$ & $-214.83\pm0.08$ & --- \\
\bottomrule
\end{tabular}
\end{center}
\caption{Total MBE-FCI/cc-pVDZ correlation energies and mean values, $\mu$, with $95\%$ confidence intervals, $\sigma$, along a given row/column (all in m$E_{\text{H}}$) for the ${^{1}}\text{A}_1$ ground state of H$_2$O ($8$e,$23$o). Expansions that failed to terminate due to insufficient screening were manually terminated at order $k=14$ (marked by ${^{\ast}}$ in the Table).}
\label{screening_table}
\end{table}
In Table \ref{screening_table}, results using any of the three types of expansions under investigation are presented for various screening thresholds (varying only the relaxation factor, $a$, of Eq. \ref{threshold}), all using the screening protocol defined by Eq. \ref{screen_cond}. As is clear from the numbers, the FCI correlation energy of $-214.80$ m$E_{\text{H}}$ is recovered to within $<0.1$ kJ/mol for all three expansion types, albeit only with screening activated. In fact, employing a static tight screening threshold throughout the expansion ($a=1.0$) or even deactivating ($a=0.0$) any screening altogether not only leads to slow convergences, and hence an increase in the cost of the total calculation, but even apparent divergences in the presence of a base model. To understand the root of this counterintuitive phenomenon---that the performance of the MBE-FCI method might deteriorate upon tightening the threshold, that is, the inclusion of an increasingly large number of subcalculations---the evolution of the absolute magnitude of the largest increment at each order is depicted in Figure \ref{inc_max_h2o_dz_relax_fig}. 

\begin{figure}[ht]
\begin{center}
\includegraphics[width=\textwidth]{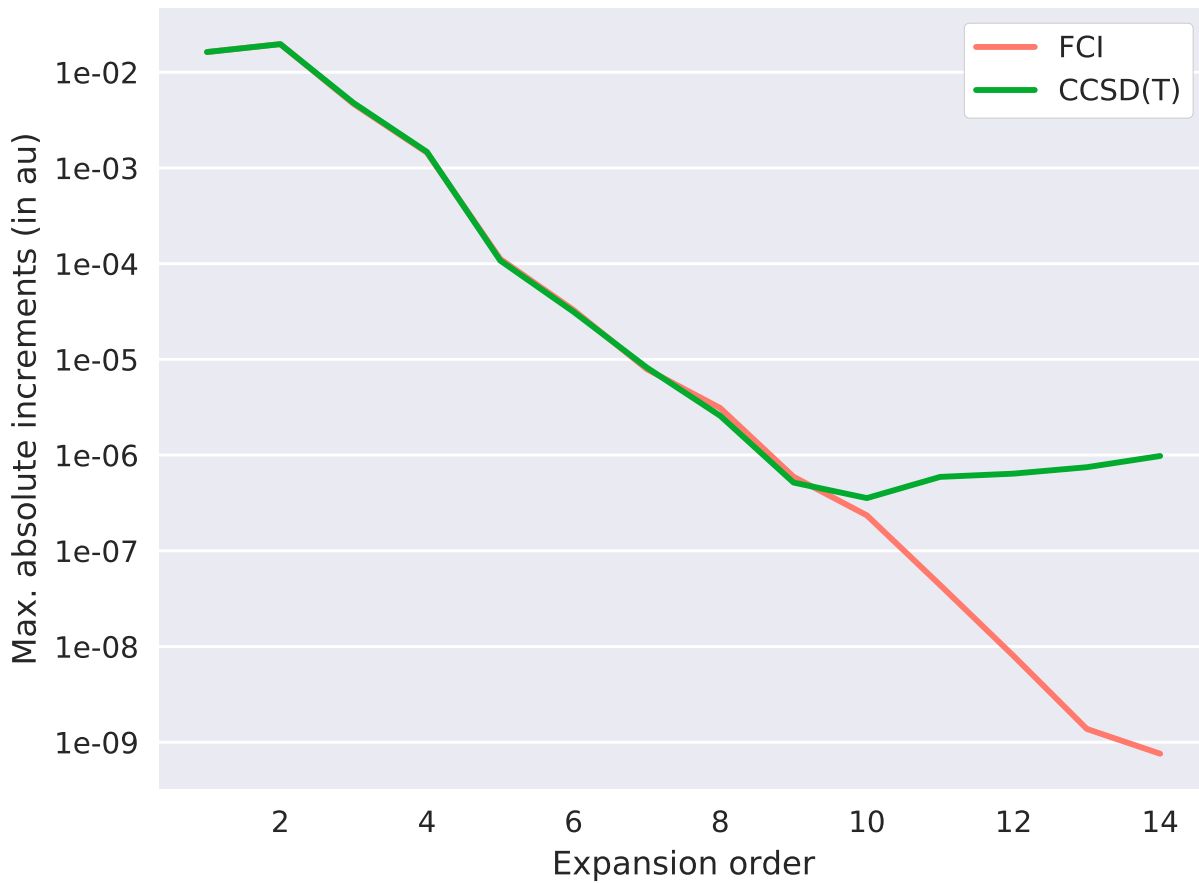}
\caption{Evolution of the largest absolute increment in MBE-FCI/cc-pVDZ and MBE-CCSD(T)/cc-pVDZ expansions ($a = 1.0$) for the ${^{1}}\text{A}_1$ ground state of H$_2$O ($8$e,$23$o).}
\label{h2o_fci_ccsdpt_diff_dz_fig}
\end{center}
\end{figure}

From the results with $a\leq1.0$ and $a>1.0$, respectively, we notice how the convergence profile of MBE-FCI changes when used in conjunction with a screening protocol. In particular, when using an intermediate base model, the introduction of which otherwise accelerates the time-to-solution, cf. Figure \ref{base_model_fig}, the convergence rate of this and the main FCI model must match for Eq. \ref{mbe_base_eq} to be useful and reliable. However, even when this holds true, any pragmatic choice of convergence threshold for the individual CC and CASCI increment calculations will ultimately influence the convergence, as, for instance, the precision of a CASCI calculation converged to within a given energy-based threshold is generally higher than a corresponding CC calculation using the same defaults. This is obvious from Figure \ref{h2o_fci_ccsdpt_diff_dz_fig}, in which the base model-free MBE-FCI curve of Figure \ref{inc_max_h2o_dz_relax_fig} ($a = 1.0$) is compared to a corresponding MBE-CCSD(T) curve using the same threshold. As discussed in the text below Eq. \ref{threshold}, we herein converge all CAS calculations below $T_{\text{init}} = \num{1.0e-10}$ $E_{\text{H}}$, and while a tighter value would serve to move the artificial minima in Figures \ref{inc_max_h2o_dz_relax_fig} and \ref{h2o_fci_ccsdpt_diff_dz_fig} to higher orders, these cannot be removed altogether from a base model-aided MBE-FCI expansion that makes use of Eq. \ref{mbe_base_eq}. Also, tightening the threshold even further will increase the cost of every single subcalculation dramatically. In the absence of a base model, however, this sort of `numerical noise' is trivially not an issue, whenever a reasonable tight $T_{\text{init}}$-value is chosen upon, as the energy summation is performed using the original Eq. \ref{mbe_eq} rather then Eq. \ref{mbe_base_eq}.
\begin{table}[ht]
\begin{center}
\begin{tabular}{lc|ccc|c}
\toprule
\multicolumn{1}{c}{Threshold} & \multicolumn{1}{c|}{Condition} & \multicolumn{1}{c}{{\textbf{b:None}}} & \multicolumn{1}{c}{{\textbf{b:CCSD}}} & \multicolumn{1}{c|}{{\textbf{b:CCSD(T)}}} & \multicolumn{1}{c}{$\mu\pm\sigma$} \\
\midrule\midrule
\multicolumn{6}{c}{cc-pVDZ} \\
\midrule
\multirow{2}{*}{$a = 10.0$} & Eq. \ref{screen_cond} & $-214.82$ & $-214.81$ & $-214.80$ & $-214.81\pm0.02$ \\
& Eq. \ref{screen_cond_old} & $-214.76$ & $-214.82$ & $-214.80$ & $-214.79\pm0.08$ \\
\multirow{2}{*}{$a = 5.0$} & Eq. \ref{screen_cond} & $-214.80$ & $-214.79$ & $-214.80$ & $-214.80\pm0.01$ \\
& Eq. \ref{screen_cond_old} & $-214.83$ & $-214.83$ & $-214.80$ & $-214.82\pm0.04$ \\
\midrule
\multicolumn{6}{c}{cc-pVTZ} \\
\midrule
\multirow{2}{*}{$a = 10.0$} & Eq. \ref{screen_cond} & $-275.51$ & $-275.39$ & $-275.30$ & $-275.40\pm0.26$ \\
& Eq. \ref{screen_cond_old} & $-274.91$ & $-275.44$ & $-275.31$ & $-275.22\pm0.69$ \\
\multirow{2}{*}{$a = 5.0$} & Eq. \ref{screen_cond} & $-275.47$ & $-275.46$ & $-275.40$ & $-275.44\pm0.09$ \\
& Eq. \ref{screen_cond_old} & $-275.22$ & $-275.36$ & $-275.31$ & $-275.30\pm0.18$ \\
\bottomrule
\end{tabular}
\end{center}
\caption{Total MBE-FCI/cc-pV$X$Z correlation energies and mean values, $\mu$, with $95\%$ confidence intervals, $\sigma$, along a given row (all in m$E_{\text{H}}$) for the ${^{1}}\text{A}_1$ ground state of H$_2$O (DZ: $8$e,$23$o; TZ: $8$e,$57$o) using Eqs. \ref{screen_cond} or \ref{screen_cond_old}  as the screening condition.}
\label{protocol_table}
\end{table}
Having established the need for an effective screening protocol, cc-pVDZ and cc-pVTZ results using either Eqs. \ref{screen_cond} or \ref{screen_cond_old} for this purpose are reported in Table \ref{protocol_table} such that these two protocols may be directly compared to one another. As discussed towards the end of Section \ref{mbe_fci_subsection}, the present protocol (Eq. \ref{screen_cond}) might be classified as conservative and the original (Eq. \ref{screen_cond_old}) as aggressive, in the sense that the condition formulated by the former of these two is stricter, yet slightly more rigorous than that formulated by the latter. That the use of Eq. \ref{screen_cond} also leads to more uniform results in practice is reflected in the statistical results of Table \ref{protocol_table}, e.g., the significantly larger confidence intervals resulting from the use of Eq. \ref{screen_cond_old}. Thus, based on the results of Tables \ref{screening_table} and \ref{protocol_table}, we will henceforth only report results obtained using the new screening protocol in Eq. \ref{screen_cond}. Furthermore, will we restrict ourselves to consider only results which have been obtained using any of the two discussed CC base models.

\subsection{Comparisons to Existing Schemes}\label{comparison_subsection}

In order to properly assess the performance of the MBE-FCI method, it must be compared to some of the existing methods discussed in Section \ref{intro_section}, which too aim at yielding FCI-level total energies. In addition, we compare results to those of the CCSDTQ model. Despite having a different formal limit from all of the other methods that we will compare with, CCSDTQ is the most advanced CC variant routinely applied as a high-accuracy approximation to FCI for systems dominated by weak electron correlation~\cite{shavitt_bartlett_cc_book,mest,ncc} and hence capable of yielding adequate near-exact reference values for such systems.

Among the previously discussed methods, comparisons will be made to the selected CI schemes mentioned earlier, in particular the SHCI method~\cite{holmes_umrigar_heat_bath_fock_space_jctc_2016,holmes_umrigar_heat_bath_ci_jctc_2016,sharma_umrigar_heat_bath_ci_jctc_2017,holmes_sharma_heat_bath_ci_excited_states_jcp_2017}, which partitions the complete set of determinants that may be generated from the HF reference into two subspaces, one which is treated variationally and a second which is treated by semistochastic second-order perturbation theory. The extent of each of these is determined by a dedicated energy threshold, and resulting electronic energies are returned with associated statistical error bars due to the stochastic sampling (cf. the Supporting Information for further discussions related to these uncertainties). As recently discussed in the context of applying this type of perturbative correction to a zeroth-order DMRG wave function~\cite{guo_chan_pdmrg_jctc_2018}, it is important to note that the variational CI energy obtained from the first of these two subspaces alone, when sampled by means of any flavour of selected CI, is rather mediocre and that the seemingly remarkable accuracy of the second-order correction stands in stark contrast to a corresponding application within traditional multireference theory. Also, whenever the threshold governing the size of the variational wave function in SHCI is so large that the subspace shrinks to encompass only the HF determinant, the perturbative correction will yield that of second-order Epstein-Nesbet perturbation theory~\cite{epstein_phys_rev_1926,nesbet_proc_1955}, the use of which is generally discouraged in favour of more traditional M\o ller-Plesset perturbation theory~\cite{mp2_phys_rev_1934}. However, Ref. \citenum{guo_chan_pdmrg_jctc_2018} argued the case that the key to the success of perturbative corrections on top of selected CI methods is related to a balance between different orbital correlations in the zeroth-order wave function, unlike in traditional multireference methods, and its use generally enables impressively fast near-exact calculations to take place even on modest hardware resources. In the present context, our choice of parameters has been motivated by the defaults and numerical results of Ref. \citenum{holmes_umrigar_heat_bath_ci_jctc_2016}, but we note that tighter thresholds alongside an extrapolation procedure may be used to achieve greater accuracy for small molecular systems dominated by weak correlation similar to those of the present study~\bibnote{Private communication with Cyrus Umrigar of Cornell University.}.

Furthermore, comparisons will be made to FCI quantum Monte Carlo in its initiator adaption~\cite{booth_alavi_fciqmc_jcp_2009,cleland_booth_alavi_jcp_2010} ($i$-FCIQMC) as well as so-called projector CI~\cite{zhang_evangelista_projector_ci_jctc_2016} (PCI). In the former of these two methods, a variational wave function is converged upon repeated stochastic application of a projection operator onto some initial state, usually a restricted HF reference, while the latter---as a deterministic realization of the same type of algorithm---combines projection onto the ground state with a path-filtering truncation scheme. Very recently, the application of Epstein-Nesbet perturbation theory has also been proposed as a means to correct for potential inadequacies in $i$-FCIQMC wave functions~\cite{blunt_fciqmc_jcp_2018}, but we stress here that all $i$-FCIQMC results reported in the present work are uncorrected. Finally, MBE-FCI will be compared to results obtained using the iFCI method by Zimmerman~\cite{zimmerman_ifci_jcp_2017_1,zimmerman_ifci_jcp_2017_2,zimmerman_ifci_jpca_2017}, which too is an MBE-based approach aimed at the FCI limit, but one in which the energy is expanded in a basis of occupied MOs and where a truncation of the virtual MO space in each CASCI subcalculation is enforced (and required).

\subsubsection{Closed-Shell Examples}\label{closed_shell_subsubsection}

In Table \ref{h2o_table}, we summarize results for H$_2$O by comparing these to available reference results of the present work as well as from the literature. In the cc-pVDZ basis set, the CCSDTQ result differs by a mere $0.01$ m$E_{\text{H}}$ from the FCI reference, a difference which might be expected to increase only slightly in the larger cc-pVTZ and cc-pVQZ basis sets, in which no FCI reference values exist. From Table \ref{h2o_table}, we recognize that MBE-FCI, aided by any of the two base models used in the present context, coincides with FCI in the cc-pVDZ basis, while showing only marginal deviations from a best estimate of the exact result within the larger triple- and quadruple-$\zeta$ basis sets. Importantly, whereas no direct error may be attached to any single MBE-FCI result, as in, e.g., SHCI, the use of different base models for any given choice of threshold returns a set of calculations that should in principle agree to within some uncertainty. For instance, the confidence intervals reported in Tables \ref{screening_table} and \ref{protocol_table} reflect this variance in the results, which diminishes upon tightening the threshold (keeping $a>1.0$ when using base models to ensure convergence, cf. the discussion above). From the results in Table \ref{h2o_table}, we observe a slight increase in this uncertainty upon increasing the basis set cardinal number, an observation which reflects the corresponding enlargement of the Hilbert space.
\begin{table}[ht]
\begin{center}
\begin{tabular}{l|cc|cccc}
\toprule
\multicolumn{1}{c|}{Threshold} & \multicolumn{1}{c}{{\textbf{b:CCSD}}} & \multicolumn{1}{c|}{{\textbf{b:CCSD(T)}}} & \multicolumn{1}{c}{iFCI\textsuperscript{\emph{a}}} & \multicolumn{1}{c}{SHCI} & \multicolumn{1}{c}{CCSDTQ} & \multicolumn{1}{c}{FCI} \\
\midrule\midrule
\multicolumn{7}{c}{cc-pVDZ} \\
\midrule
$a = 10.0$ & $-214.81$ & $-214.80$ & \multirow{2}{*}{$-213.80$} & \multirow{2}{*}{$-214.84\pm0.02$} & \multirow{2}{*}{$-214.79$} & \multirow{2}{*}{$-214.80$} \\
$a = 5.0$ & $-214.79$ & $-214.80$ & & & & \\
\midrule
\multicolumn{7}{c}{cc-pVTZ} \\
\midrule
$a = 10.0$ & $-275.39$ & $-275.30$ & \multirow{2}{*}{$-274.50$} & \multirow{2}{*}{$-275.71\pm0.07$} & \multirow{2}{*}{$-275.36$} & \multirow{2}{*}{N/A} \\
$a = 5.0$ & $-275.46$ & $-275.40$ & & & & \\
\midrule
\multicolumn{7}{c}{cc-pVQZ} \\
\midrule
$a = 10.0$ & $-297.05$ & $-295.77$ & \multirow{2}{*}{N/A} & \multirow{2}{*}{$-295.96\pm0.09$} & \multirow{2}{*}{$-295.29$} & \multirow{2}{*}{N/A} \\
$a = 5.0$ & $-295.20$ & $-295.15$ & & & & \\
\bottomrule
\end{tabular}
\\
\textsuperscript{\emph{a}} iFCI ($\zeta = \num{1e-6.5}$) results from Table V of Ref. \citenum{zimmerman_ifci_jcp_2017_1}.
\end{center}
\caption{Total MBE-FCI/cc-pV$X$Z correlation energies (in m$E_{\text{H}}$) for the ${^{1}}\text{A}_1$ ground state of H$_2$O (DZ: $8$e,$23$o; TZ: $8$e,$57$o; QZ: $8$e,$114$o). For comparison, corresponding iFCI, SHCI, CCSDTQ, and FCI results are also presented.}
\label{h2o_table}
\end{table}

As touched upon above, CCSDTQ is expected to give a correlation energy for H$_2$O only slightly smaller in magnitude than that of FCI in the cc-pVTZ basis, that is, the exact reference is expected to be close to the {\textbf{b:CCSD(T)} result of $-275.40$ m$E_{\text{H}}$. For iFCI and SHCI, we note that the former of these two underestimates the correlation energy by approximately $1.0$ m$E_{\text{H}}$ in both basis set (an error which is also influenced by the resolution-of-the-identity (RI) approximation~\cite{feyereisen_rimp2_cpl_1993} used throughout the iFCI scheme) while the latter overestimates the correlation energy, albeit by less than iFCI underestimates it in both basis sets. As noted above, the error of SHCI may be further reduced, but so may the error of iFCI and MBE-FCI by using less pragmatic thresholds, while this is trivially not true in the case of CCSDTQ. In the largest of the tested basis sets, cc-pVQZ, MBE-FCI most likely yields a marginal underestimation of the correlation energy, as is visible from the deviations from CCSDTQ of $0.2$ to $0.4$ kJ/mol, differences which are expected not to be much different from those with respect to the exact result.

Next, we investigate to what degree MBE-FCI lacks size consistency by performing calculations for the same type of test system as was recently used by Zhang and Evangelista in Ref. \citenum{zhang_evangelista_projector_ci_jctc_2016}, namely a Be--He monomer and the corresponding noninteracting dimer, both arranged in $C_{\infty\text{v}}$ symmetry. The same mixed basis set is used as in Ref. \citenum{zhang_evangelista_projector_ci_jctc_2016} (Be: cc-pVDZ, He: STO-3G), and no core orbitals are kept frozen in neither the monomer nor the dimer calculation. The MBE-FCI results are collected in Table \ref{be_he_table}.
\begin{table}[ht]
\begin{center}
\begin{tabular}{l|cc|ccc}
\toprule
\multicolumn{1}{c|}{Threshold} & \multicolumn{1}{c}{{\textbf{b:CCSD}}} & \multicolumn{1}{c|}{{\textbf{b:CCSD(T)}}} & \multicolumn{1}{c}{PCI\textsuperscript{\emph{a}}} & \multicolumn{1}{c}{SHCI} & \multicolumn{1}{c}{FCI} \\
\midrule\midrule
\multicolumn{6}{c}{Be--He} \\
\midrule
$a = 10.0$ & $-46.4207$ & $-46.4208$ & \multirow{2}{*}{$-46.420$} & \multirow{2}{*}{$-46.4194\pm0.0002$} & \multirow{2}{*}{$-46.4206$} \\
$a = 5.0$ & $-46.4206$ & $-46.4207$ & & & \\
\midrule
\multicolumn{6}{c}{He--Be$\cdots$Be--He} \\
\midrule
$a = 10.0$ & $-92.8415$ & $-92.8414$ & \multirow{2}{*}{$-92.838$} & \multirow{2}{*}{$-92.8379\pm0.0010$} & \multirow{2}{*}{$-92.8413$} \\
$a = 5.0$ & $-92.8408$ & $-92.8410$ & & & \\
\midrule
\multicolumn{6}{c}{Size Consistency Error} \\
\midrule
$a = 10.0$ & $0.1$ & $0.2$ & \multirow{2}{*}{$2$} & \multirow{2}{*}{$0.1-2.1$} & \multirow{2}{*}{$0.0$} \\
$a = 5.0$ & $0.4$ & $0.4$ & & & \\
\bottomrule
\end{tabular}
\\
\textsuperscript{\emph{a}} PCI ($\eta = \num{1e-06}$) results in a basis of MP2 NOs from Table 4 of Ref. \citenum{zhang_evangelista_projector_ci_jctc_2016}.
\end{center}
\caption{Total MBE-FCI/cc-pVDZ/STO-3G correlation energies (in m$E_{\text{H}}$) for the ${^{1}}\text{A}_{1}$ ground states of Be--He ($6$e,$15$o) and (Be--He)$_2$ ($12$e,$30$o) as well as the associated absolute size consistency error (in $\mu E_{\text{H}}$). For comparison, corresponding PCI, SHCI, and FCI results are also presented.}
\label{be_he_table}
\end{table}

As is obvious from the results in Table \ref{be_he_table}, the accuracy of MBE-FCI in reproducing both total as well as relative FCI energies is again high; in fact, any deviation from FCI is only visible upon increasing the number of reported significant figures with respect to Tables \ref{screening_table}--\ref{h2o_table}, despite all settings being fixed to the same default values as used everywhere else in this work (canonical occupied MOs, same CASCI convergence threshold, etc.). For comparison, SHCI and PCI results are also presented in Table \ref{be_he_table}. Both of these sets of results, which are obtained in similar delocalized NO bases, show slightly larger size consistency errors. Hence, although the present test system is artificial, these results demonstrate that errors affiliated with size inconsistency are negligible in MBE-FCI when compared to total errors with respect to FCI.

As a final closed-shell example, we direct our attention at ethylene, C$_2$H$_4$, which is another prototypical molecule dominated by weak electron correlation, albeit one slightly larger than, e.g., H$_2$O. We use the ANO-L-VDZP basis set~\cite{roos_ano} with the same contraction scheme as in Ref. \citenum{daday_booth_alavi_filippi_fciqmc_jctc_2012} in order to facilitate direct comparisons to alternative FCI-level methods. Two sets of geometries are used, namely those of Refs. \citenum{daday_booth_alavi_filippi_fciqmc_jctc_2012} and \citenum{zimmerman_ifci_jcp_2017_1}.

\begin{table}[ht]
\begin{center}
\begin{tabular}{l|cc|cccc}
\toprule
\multicolumn{1}{c|}{Threshold} & \multicolumn{1}{c}{{\textbf{b:CCSD}}} & \multicolumn{1}{c|}{{\textbf{b:CCSD(T)}}} & \multicolumn{1}{c}{iFCI\textsuperscript{\emph{a}}} & \multicolumn{1}{c}{$i$-FCIQMC\textsuperscript{\emph{b}}} & \multicolumn{1}{c}{SHCI} & \multicolumn{1}{c}{CCSDTQ} \\
\midrule\midrule
\multicolumn{7}{c}{Geometry from Ref. \citenum{zimmerman_ifci_jcp_2017_1}} \\
\midrule
$a = 10.0$ & $-305.53$ & $-305.47$ & \multirow{2}{*}{$-304.81$} & \multirow{2}{*}{N/A} & \multirow{2}{*}{$-304.69\pm0.09$} & \multirow{2}{*}{$-305.41$} \\
$a = 5.0$ & $-305.44$ & $-305.45$ & & & & \\
\midrule
\multicolumn{7}{c}{Geometry from Ref. \citenum{daday_booth_alavi_filippi_fciqmc_jctc_2012}} \\
\midrule
$a = 10.0$ & $-305.98$ & $-305.93$ & \multirow{2}{*}{N/A} & \multirow{2}{*}{$-305.0\pm0.1$} & \multirow{2}{*}{$-305.26\pm0.08$} & \multirow{2}{*}{$-305.92$} \\
$a = 5.0$ & $-306.01$ & $-305.96$ & & & & \\
\bottomrule
\end{tabular}
\\
\textsuperscript{\emph{a}} iFCI/$\zeta = \num{1e-6.5}$/$n=3$ results from Table V of Ref. \citenum{zimmerman_ifci_jcp_2017_1}.
\\
\textsuperscript{\emph{b}} $i$-FCIQMC results from Table 3 of Ref. \citenum{daday_booth_alavi_filippi_fciqmc_jctc_2012}.
\end{center}
\caption{Total MBE-FCI/ANO-L-VDZP correlation energies (in m$E_{\text{H}}$) for the ${^{1}}\text{A}_{\text{g}}$ ground state of C$_2$H$_4$ ($12$e,$46$o). For comparison, corresponding iFCI, $i$-FCIQMC, SHCI, and CCSDTQ results are also presented.}
\label{c2h4_table}
\end{table}
In Table \ref{c2h4_table}, MBE-FCI results are compared to corresponding SHCI and CCSDTQ results of the present work as well as iFCI and $i$-FCIQMC results from the literature. Again, as no FCI reference value is available, it is fair to assume that the CCSDTQ result of $-305.41$ m$E_{\text{H}}$ lies close to, although most likely slightly above the exact result. To that end, we notice that such a prediction matches the uniform MBE-FCI results, which, in turn, are below those of any of the alternative methods presented alongside MBE-FCI in Table \ref{c2h4_table}.

\subsubsection{Open-Shell Examples}\label{open_shell_subsubsection}

Having explored the performance of MBE-FCI for three closed-shell systems, we now turn to two open-shell systems, namely the methylene and oxygen molecules. The geometry for the ${^{3}}\text{B}_1$ ground state of methylene is adapted from a recently published study of singlet-triplet gaps by Yang et al.~\cite{yang_singlet_triplet_gaps_jpca_2015}, while for the oxygen molecule in its ${^{3}}\Sigma^{-}_{\text{g}}$ ground state, the experimental structure as reported by Huber and Herzberg~\cite{huber_herzberg_geo_book} is used. Like for water in Section \ref{screening_subsection}, we again use the standard cc-pV$X$Z basis sets.

\begin{table}[ht]
\begin{center}
\begin{tabular}{l|cc|ccc}
\toprule
\multicolumn{1}{c|}{Threshold} & \multicolumn{1}{c}{{\textbf{b:CCSD}}} & \multicolumn{1}{c|}{{\textbf{b:CCSD(T)}}} & \multicolumn{1}{c}{SHCI} & \multicolumn{1}{c}{CCSDTQ} & \multicolumn{1}{c}{FCI} \\
\midrule\midrule
\multicolumn{6}{c}{cc-pVDZ} \\
\midrule
$a = 10.0$ & $-120.27$ & $-120.27$ & \multirow{2}{*}{$-120.29\pm0.01$} & \multirow{2}{*}{$-120.28$} & \multirow{2}{*}{$-120.28$} \\
$a = 5.0$ & $-120.28$ & $-120.27$ & & & \\
\midrule
\multicolumn{6}{c}{cc-pVTZ} \\
\midrule
$a = 10.0$ & $-146.10$ & $-146.13$ & \multirow{2}{*}{$-146.22\pm0.04$} & \multirow{2}{*}{$-146.15$} & \multirow{2}{*}{N/A} \\
$a = 5.0$ & $-146.11$ & $-146.11$ & & & \\
\bottomrule
\end{tabular}
\end{center}
\caption{Total MBE-FCI/cc-pV$X$Z correlation energies (in m$E_{\text{H}}$) for the ${^{3}}\text{B}_1$ ground state of CH$_2$ (DZ: $6$e,$23$o; TZ: $6$e,$57$o). For comparison, corresponding ROHF-based SHCI, CCSDTQ, and FCI results are also presented.}
\label{ch2_table}
\end{table}
Before assessing the accuracy of the MBE-FCI results for CH$_2$ and O$_2$ in Tables \ref{ch2_table} and \ref{o2_table}, respectively, we initially comment on the use of CCSD and CCSD(T) as base models in the case of open-shell species. In a previous study of the performance of perturbative CC models for open-shell species by the present authors~\cite{open_shell_triples_jcp_2016,open_shell_quadruples_jcp_2016}, a discrepancy between the performance of CCSD and, notably, CCSD(T) was observed in moving from closed-shell to open-shell species, which was particularly striking for O$_2$ when described at the mean-field level by an ROHF reference determinant. However, as discussed towards the end of Section \ref{new_subsection}, any lack of rigour in the CCSD(T) model is not necessarily of concern for its application as an MBE-FCI base model as long as it remains qualitatively accurate. As is obvious from the results in Tables \ref{ch2_table} and \ref{o2_table}, this is indeed the case for both systems tested here.

\begin{table}[ht]
\begin{center}
\begin{tabular}{l|cc|ccc}
\toprule
\multicolumn{1}{c|}{Threshold} & \multicolumn{1}{c}{{\textbf{b:CCSD}}} & \multicolumn{1}{c|}{{\textbf{b:CCSD(T)}}} & \multicolumn{1}{c}{$i$-FCIQMC\textsuperscript{\emph{a}}} & \multicolumn{1}{c}{SHCI} & \multicolumn{1}{c}{CCSDTQ} \\
\midrule\midrule
$a = 10.0$ & $-379.69$ & $-379.70$ & \multirow{2}{*}{$-379.73\pm0.08$} & \multirow{2}{*}{$-379.65\pm0.10$} & \multirow{2}{*}{$-379.52$} \\
$a = 5.0$ & $-379.71$ & $-379.71$ & & & \\
\bottomrule
\end{tabular}
\\
\textsuperscript{\emph{a}} $i$-FCIQMC results from Table 1 of Ref. \citenum{cleland_booth_alavi_fciqmc_jctc_2012}.
\end{center}
\caption{Total MBE-FCI/cc-pVDZ correlation energies (in m$E_{\text{H}}$) for the ${^{3}}\Sigma^{-}_{\text{g}}$ ground state of O$_2$ ($12$e,$26$o). For comparison, corresponding ROHF-based $i$-FCIQMC, SHCI, and CCSDTQ results are also presented.}
\label{o2_table}
\end{table}
For CH$_2$ in Table \ref{ch2_table}, we notice that the MBE-FCI results coincide exactly with the FCI (and CCSDTQ) results in the cc-pVDZ basis, regardless of the level of screening, while in the larger cc-pVTZ basis, all results are centred around a correlation energy of $-146.11$ m$E_{\text{H}}$. This value is slightly above the corresponding CCSDTQ result, which leaves some doubt as to which result is closest to the exact result. In both basis sets, however, the MBE-FCI results tend to be in better agreement with FCI than the corresponding SHCI results, which, for this molecule, tend to be too large in magnitude.

Finally, for O$_2$ in Table \ref{o2_table}, the variance of the MBE-FCI results with base model and screening threshold is once again marginal, with a best estimate of the exact correlation energy of $-379.71$ m$E_{\text{H}}$. As opposed to all of the other examples of the present work, the CCSDTQ result for O$_2$ is clearly too small in magnitude, as is recognized not only from a comparison with MBE-FCI, but also the $i$-FCIQMC result of Ref. \citenum{cleland_booth_alavi_fciqmc_jctc_2012}, which matches our best estimate, albeit with significantly larger uncertainties attached to it. Error bars of the same magnitude are also attached to the SHCI result in Table \ref{o2_table}, but here the result lies somewhere in-between CCSDTQ and MBE-FCI, i.e., it is slightly too small in magnitude.

%
%%%%%%%%%%%%%%%%%%%%%%%%%%%%%%%%%%%%%%%%%%%%%%%%%%%%%%%%%%%%%%%%%%%
%                                                                    		CONCLUSIONS AND OUTLOOK
%%%%%%%%%%%%%%%%%%%%%%%%%%%%%%%%%%%%%%%%%%%%%%%%%%%%%%%%%%%%%%%%%%%
%

\section{Summary and Conclusions}\label{conclusion_section}

Leveraged by technical advances as well as a large volume of novel techniques, the past decade has seen the ideas behind selected CI gain renewed momentum through the introduction of various incarnations which all seek to sample the Hilbert space in inventive and focused rather than complete manners. To that end, all of these proposed algorithms share in common a diversion from the norm of approximating the FCI wave function in terms of a fixed, yet reduced set of variable parameters by instead striving to compress its most important components to the largest extent possible. As elusive as it might sound, FCI-level calculations for diverse types and sizes of molecular systems are becoming increasingly feasible these days, and the particular niche of theoretical chemistry to which such work relates is hence gathering a proportionally increased amount of attention as a direct consequence.

In the present work, a number of advancements of one such recent addition to this growing body of near-exact methods, MBE-FCI, have been detailed and numerically tested for a sample of weakly correlated closed- and open-shell species. In MBE-FCI, the exact correlation energy within a given one-electron basis set is decomposed by means of a many-body expansion in the individual virtual orbitals of a preceding mean-field calculation. Facilitated by a screening protocol to ensure convergence onto the FCI target, and aided as well as accelerated by intermediate CC base models, results of sub-kJ/mol accuracy have been obtained on commodity hardware for the molecules H$_2$O, C$_2$H$_4$, CH$_2$, and O$_2$ in standard correlation-consistent basis sets of double-, triple, and quadruple-$\zeta$ quality. The overall reliability of MBE-FCI has been assessed and its precision validated through numerical comparisons against both exact and similar high-accuracy reference results, and a numerical investigation of the (negligible) lack of size consistency in the method has furthermore been presented.

Despite offering a significantly more expensive alternative than selected CI counterparts such as, e.g., SHCI, the embarrassingly parallel nature of the MBE-FCI algorithm holds promise of providing a tractable route towards FCI-level results for larger basis sets as well, as will be explored numerically in future studies. However, given the formulation of MBE-FCI on top of an underlying HF reference, a bias towards the weakly correlated regime prohibits the controllable application of the method to systems of interest which are dominated more by static than dynamic correlation. In order to relieve this dependence, the following part of the present series will be devoted to an intuitive extension of MBE-FCI in terms of multireference expansion references, which, in addition to the general dividends resulting from the performance enhancements of the method outlined in the present work, will enable us to target also the considerably more demanding regime of strongly correlated systems.

%
%%%%%%%%%%%%%%%%%%%%%%%%%%%%%%%%%%%%%%%%%%%%%%%%%%%%%%%%%%%%%%%%%%%
%                                                                    			     ACKNOWLEDGMENT
%%%%%%%%%%%%%%%%%%%%%%%%%%%%%%%%%%%%%%%%%%%%%%%%%%%%%%%%%%%%%%%%%%%
%
\section*{Acknowledgments}

J. J. E. wishes to thank Sandeep Sharma of University of Colorado Boulder for discussions on the {\textsc{dice}} code and Qiming Sun of California Institute of Technology for his general work on the {\textsc{pyscf}} code. J. J. E. is grateful to the Alexander von Humboldt Foundation for financial support.

%
%%%%%%%%%%%%%%%%%%%%%%%%%%%%%%%%%%%%%%%%%%%%%%%%%%%%%%%%%%%%%%%%%%%
%                                                                    			     Supporting Information
%%%%%%%%%%%%%%%%%%%%%%%%%%%%%%%%%%%%%%%%%%%%%%%%%%%%%%%%%%%%%%%%%%%
%
\section*{Supporting Information}

Details on convergence and cut-off thresholds, sampling spaces, etc., for the CCSDTQ and SHCI reference calculations are collected in the Supporting Information, as are Cartesian coordinates and reference HF, CCSD, and CCSD(T) energies for all of the geometries used herein.

\newpage

\providecommand*\mcitethebibliography{\thebibliography}
\csname @ifundefined\endcsname{endmcitethebibliography}
  {\let\endmcitethebibliography\endthebibliography}{}

\end{document}